\documentstyle[nato,epsfig]{crckapb}
\def\i{\begin{equation}}
\def\e{\end{equation}}

\begin{opening}
\title{ POLARIZATION OF THE MICROWAVE BACKGROUND:\protect\\ THEORETICAL FRAMEWORK}
\author{ALESSANDRO MELCHIORRI}
\author{NICOLA VITTORIO}
\institute{ Dipartimento di Fisica \\  Universit\`a "Tor Vergata", Roma, Italy}
\end{opening}
\begin{document}

\begin{abstract}

We present a brief review of the polarization properties of the cosmic
 microwave background in dark matter models for structure formation.
Quite independently of the model parameters, the polarization level is expected
 to be $\sim 10 \%$ of the anisotropy signal at angular scales $\le 1^o$.
Detections of polarization at larger angular scales would provide a
strong evidence in favour of an early reionization of the intergalactic 
medium.
\end{abstract}

\section{ Introduction: some historical remarks}

Most of the early theoretical work on the polarization of the Cosmic Microwave
Background (CMB) was focused, after Rees pioneering work [1], on anisotropic 
cosmological models [2,3,4,5,6,7,8,9].
The degree of Faraday rotation expected in 
 the presence of an universal magnetic field  and 
the use of polarization measurements to constraint the amplitude
of such field were also considered [14,33]. 
More recently, it has been shown that even in isotropic cosmological models the
anisotropic component of the CMB is polarized [10]. Detailed numerical 
predictions have been made for dark matter dominated models
with adiabatic fluctuations [see e.g. 11,12,13], with and
without an early reionization of the intergalactic medium [15].These calculations
have shown that the level of polarization can be $10 $ percent of the anisotropy
signal.
After the COBE/DMR result [16], new attention has been dedicated
to the tensor modes of metric fluctuations, which also produces anisotropy
on large angular scales [17,18,19,20,21,22,23]. The polarization
due to a background of primordial gravitational waves has been widely
discussed [24,25,26,27,28].
For describing the statistic of the polarization field was also 
introduced the polarization - anisotropy correlation function [29,30], 
while other authors [31,32] have shown that neglecting polarization
yields a theoretical overestimate of the anisotropy at small angular scale.

\
From the experimental side, in spite of a continuous increment in the experimental sensitivity, no
polarization was found and only upper limits were given 
[34,35,36,37], with the best upper limit to date of $\sim 25 \mu K$ from
the Saskatoon experiment [38].
As we show in Section 5, the level of CMB polarization expected is in most of the models at least a
factor $10$ below this limit, so is not clear if the present sensitivity of the CMB 
experiments is sufficient to detect polarization.
However, in view of forthcoming  high sensitivity new experiments,  
it is of interest to discuss the general properties of the polarization pattern and its dependence
to the various cosmological parameters.
So, the aim of this work is to review the basics steps behind the theoretical
calculation of CMB polarization. The plan of the paper is as follows. In 
Section $2$ we briefly review the definition of the Stokes
parameters and their variations in a Thomson scattering.
In Section $3$ and $4$ we write the set of equations necessary to describe anisotropy
and polarization of the CMB. In Section $5$ we review some of the results 
obtained by numerically integrate those equations. 
Finally, in Section $6$, we summarize the main findings.

\section{ An Elementary Description of the Polarization of Light}

\begin{figure}
\centerline{\epsfig{file=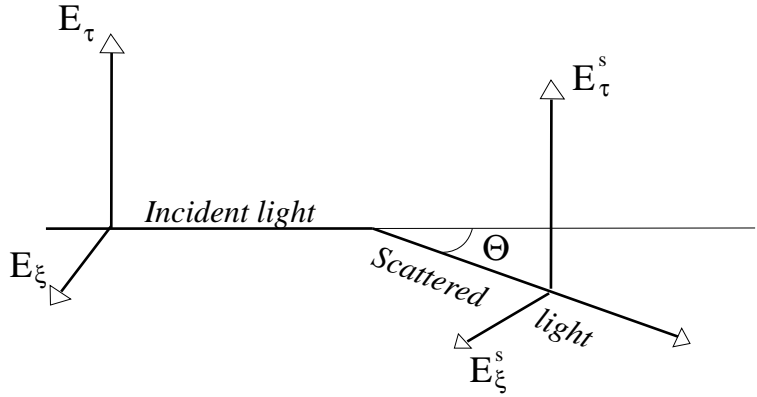,width=2.4in,angle=-90}}
\caption{Thomson scattering of a photon by an electron.}
\end{figure}

For an elliptically polarized wave, the components of the electric field along two orthogonal 
directions, $\xi$ and $\tau$ say, can be written as :

\i 
\cases{ E_{\xi} = E_{\xi}^0 \sin(\omega t-\epsilon_1)\cr
 E_\tau=E_\tau^0 \sin(\omega t-\epsilon_2)\cr}
\e

\noindent where $E^0_{\xi,\tau}$ and $\epsilon_{1,2}$ are constants.
The polarization of the radiation field is conveniently described in 
terms of the Stokes parameters :

\i 
\cases{ I = {E_{\xi}^0}^2 + {E_{\tau}^0}^2 \equiv I_\xi + I_\tau\cr
Q = {E_{\xi}^0}^2 - {E_{\tau}^0}^2 \equiv I_\xi - I_\tau\cr
U = 2 E_{\xi}^0 E_\tau^0 \cos[\epsilon_1 - \epsilon_2]\cr
V= 2 E_{\xi}^0 E_\tau^0 \sin[\epsilon_1 - \epsilon_2]\cr}
\e 

The parameter $I$ is proportional to the intensity of the wave (we
omit the proportionality factor) $V$ measures the ratio of the principal
axes of the polarization ellipse while $Q$ or $U$ measures the orientation of
the ellipse relative to the $\xi$ axes. In general $I^2 \ge Q^2 + U^2 + V^2$, the
equality holding for an elliptically polarized wave.

\

A clockwise rotation by an angle $\Xi$ of the $\xi - \tau$ axes in the
polarization plane leaves unchanged the $I$ and $V$ parameters and it is
equivalent to apply the operator:

\i {\bf \hat{L}}(\Xi) = \pmatrix{\cos^2\Xi&\sin^2\Xi&{\sin 2\Xi\over 2}&0\cr
\sin^2\Xi&\cos^2\Xi&-{\sin 2\Xi\over 2}&0\cr
-\sin2\Xi&\sin2\Xi&\cos2\Xi&0\cr
0&0&0&1\cr}\e

\noindent to the vector ${\vec I} \equiv (I_\xi,I_\tau,U,V)$. 

In the Thomson scattering, the light scattered in a direction making an
angle $\Theta$ with the direction of incidence (see Figure 1) is

\i \cases{ E_{\xi}^s = \sqrt{{3\over 2}\sigma_T } E_{\xi}^0\cos\Theta\sin(\omega t-\epsilon_1)\cr
E_\tau^s = \sqrt{{3\over 2}\sigma_T } E_\tau^0\sin(\omega t-\epsilon_2)\cr}\e
 
In analogy with the equation ($2$) the Stokes parameters of the scattered
light are:
\i
\cases{I_\xi^s = (3/2) \sigma_T I_\xi \cos^2\Theta\cr
I_\tau^s = (3/2) \sigma_T I_\tau\cr
U^s = (3/2) \sigma_T U \cos\Theta\cr
V^s = (3/2) \sigma_T V \cos\Theta\cr}
\e

\noindent or, in matrix form, ${\vec I}^s = \sigma_T {\hat{R} \times \vec I}$, where 

\i  {\hat{R}}= {3 \over 2} \pmatrix {\cos^2\Theta&0&0&0\cr
0&1&0&0\cr
0&0&\cos\Theta&0\cr
0&0&0&\cos\Theta\cr}\e

\

\begin{figure}
\centerline{\epsfig{file=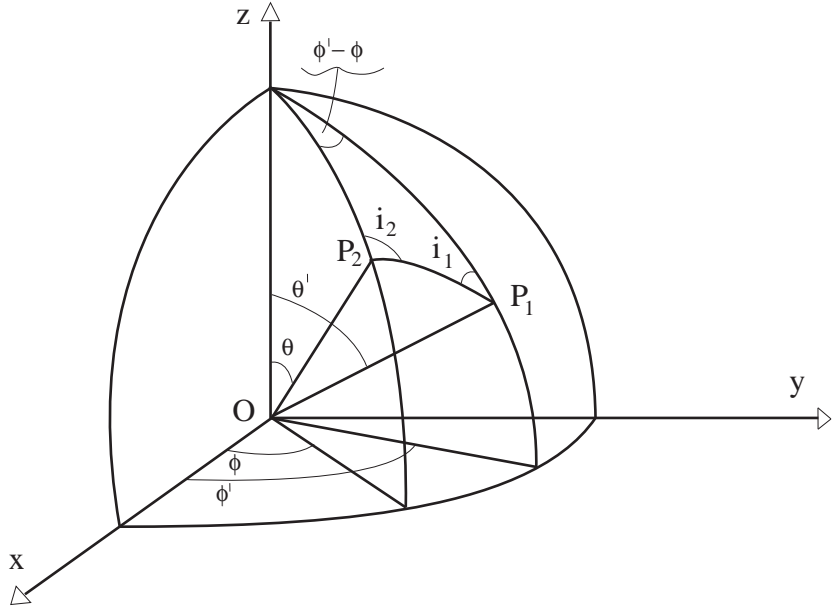,width=3in,angle=-90}}
\caption{Coordinate system needed to describe Thomson
 scattering in the Lab frame.}
\end{figure}

\

In order to study the variations of Stokes parameters in the Lab frame 
(see Figure 2) we have to :

\begin{enumerate}

\item apply the transformation ${\hat{L}}(-i_1)$  to $\vec I$,
 where $i_1$ is the angle
between the meridian and the scattering planes. In this way we obtain the Stokes parameters of
the incident light in the frame of Figure 1.

\item apply ${\hat{R}}$ to these parameters in order to
 obtain the Stokes parameters
of the scattered light, again in the frame of Figure 1.

\item apply the transformation $ {\hat{L}}(\pi-i_2) $,
 where $i_2$ represents the angle
between the plane $OP_2Z$ and $OP_1 P_2$.
 In this way we are back to the Lab frame.

\end{enumerate}

Thus, the radiation scattered in the $(\theta, \phi)$ direction, relative to the Lab frame,
can be written as [39]:

\i {\vec I}^s (\theta, \phi)=
{1\over 4
\pi} \int_{4\pi} [ {\hat{P}}(\theta,\phi;\theta',\phi')\times  {\vec I}(\theta',\phi')]d \Omega'\e

\noindent where  

\i {\hat{P}} = {\hat{Q}} \times \bigl[ {\hat{P}^0}(\mu,\mu')
+\sqrt{(1-\mu^2)}
\sqrt{(1-\mu'^2)} {\hat{P}^1} (\mu,\phi,\mu',\phi')+ {\hat{P}^2}
(\mu,\phi,\mu',\phi')\bigr]\e

\i {\hat{Q}}= \pmatrix{1&0&0&0\cr
0&1&0&0\cr
0&0&2&0\cr
0&0&0&2\cr},\e

\

\i{\hat P}^0 = {3\over4}\pmatrix{2(1-\mu^2)(1-\mu'^2)+\mu^2\mu'^2&\mu^2&0&0\cr
\mu'^2&1&0&0\cr
0&0&0&0\cr
0&0&0&\mu\mu'\cr},\e

\

\i {\hat P}^1= {3\over
4}\pmatrix{4\mu\mu'\cos(\phi-\phi')&0&2\mu\sin(\phi'-\phi)&0\cr
0&0&0&0\cr
-2\mu'\sin(\phi-\phi')&0&\cos(\phi-\phi')&0\cr
0&0&0&\cos(\phi-\phi')\cr},\e

\

\i{\hat P}^2= {3\over 4}\pmatrix{\mu^2\mu'^2\cos2(\phi'-\phi)&
-\mu^2\cos2(\phi'-\phi)&\mu^2\mu'\sin2(\phi'-\phi)&0\cr
-\mu'^2\cos2(\phi'-\phi)&\cos2(\phi'-\phi)&-\mu'\sin2(\phi'-\phi)&0\cr
-\mu'^2\mu\sin2(\phi'-\phi)&\mu\sin2(\phi'-\phi)&\mu'\mu\cos2(\phi'-\phi)&0\cr
0&0&0&0\cr},\e

\noindent and  $\mu$ and $\mu'$ are defined as  $\cos \theta$ and  $\cos \theta'$,
respectively.

\section{ The Boltzmann Transfer Equation for Polarized Light}

In order to study anisotropy and polarization of the CMB we need to write down
the transfer equation for the Stokes parameters. We restrict ourselves to
isotropic universes where, to zero-th order, anisotropy and polarization vanish.
The perturbations to the Stokes parameters and to the other relevant quantities
(see section 3.3) are written in the synchronous gauge formalism. Following
Peebles [40,41] we introduce the fractional fluctuations of the Stokes
parameters  as follows:

\i \pmatrix{I\cr
Q\cr
U\cr
V\cr}= 
{{{\rho_\gamma}(t)}\over{4\pi}}\pmatrix{1+\iota \cr
q\cr
u\cr 
v\cr}\e

\noindent where ($\iota$,$q$,$u$,$v$) are functions of the observer position $\vec x$,
of the line of sight observation $\hat{\gamma} \equiv (\gamma_1,\gamma_2,\gamma_3)$ and of the cosmic time $t$.
To first order the transfer equations becomes 

\i {{\partial} \over {\partial t}} \pmatrix{\iota\cr
q\cr
u\cr
v\cr}
+ {\gamma_{\alpha} \over a }
{{\partial} \over {\partial x^{\alpha}}} \pmatrix{\iota\cr
q\cr
u\cr
v\cr} 
+\pmatrix{y\cr 0\cr 0\cr 0\cr}  
= {\sigma_T n_e \biggl[ \pmatrix{\iota^s\cr
q^s\cr
u^s\cr
v^s\cr} - \pmatrix{\iota\cr
q\cr
u\cr
v\cr} \biggr]} \e

\

\noindent where  $y=- {2 {\dot h}_{\alpha \beta} \gamma_\alpha \gamma_\beta}$ is
 the term containing the linear perturbation to the
metric tensor, ($\iota^s$,$q^s$,$u^s$,$v^s$) are evaluated in the comoving
frame and refer to the radiation scattered in the $\vec \gamma$ direction, and
$a$ is the scale factor.
In order to avoid spatial dependence it is convenient to work in Fourier space.
We choose for each $k$ mode a reference system
with the $z$ axis parallel to $\vec k$, in order to achieve an azimuthal
symmetry.

\subsection{Scalar modes}

For scalar modes, the only non vanishing components of the perturbed metric tensor are the diagonal ones :
$h_{11}$, $h_{22}$, $h_{33}$ ($h_{00} \equiv 0$ because of the chosen gauge). Thus, the gravitational term in  equation (14) has
the form: $y = (1-3\mu^{2}){\dot h_{33}}-(1-\mu{^2}){\dot h}$, and  each Fourier mode is independent of the azimuthal 
angle $\phi$.
After integrating over this angle equation (7) it can be proved
 that $\hat{P^1}$ and $\hat{P^2}$ give no contribution. Therefore we can assume $U=0$
in this case. Also the equation for $V$  is decoupled from the others: if $V$ vanishes at the
beginning, it also vanishes afterwards. Therefore only the  perturbations
$\iota$ and $q$ of the $I$ and $Q$ parameters are of interest.
Their evolution is described by the following transfer equation [10,11,12,13]:

\i {{\partial} \over {\partial t}} \pmatrix{\iota\cr q\cr} + {i k \mu\over a}
\pmatrix{\iota\cr q\cr} - \pmatrix {y\cr0\cr}= 
\sigma_T n_e \biggl( 
  \int_{-1}^1 \hat{M_S}(\mu,\mu') \pmatrix{\iota'\cr
q'\cr}d\mu'
- \pmatrix{{\iota + 4\mu v_b} \cr q \cr}\biggr)\e 

\

\noindent where $ y = (1-3\mu^{2}){\dot h_{33}}-(1-\mu{^2}){\dot h} $ is 
the term taking into account the effects of gravitational potential,
 and where the $2 \times 2$  matrix $\hat{M_S}$ is composed by the
first two rows and columns of the matrix $\hat{P^0}$ in the $(I,Q,U,V)$ basis:

\i \hat{M_S}(\mu,\mu') = {3\over16}\pmatrix{3-\mu'^2-\mu^2+3\mu^2\mu'^2&1-\mu'^2-
3\mu^2(1-\mu'^2)\cr
1-3\mu'^2-\mu^2+3\mu^2\mu'^2&3-3\mu'^2-3\mu^2(1-\mu'^2)\cr}\e

The angular dependence in equation (15) can be eliminated by expanding $\iota$
and $q$ in Legendre polynomials :

\i \iota= \sum_\ell (\sigma^k_{2\ell} (t) P_{2\ell}(\mu) + i \sigma^k_{2\ell+1}
(t) P_{2\ell+1} (\mu))\e

\i q= \sum_\ell (\eta^k_{2\ell} (t) P_{2\ell}(\mu) + i \eta^k_{2\ell+1} (t)
P_{2\ell+1} (\mu))\e

\noindent Because of the orthogonality  of the Legendre polynomials, equation (15) becomes:

\i 
\cases{ {{\partial \iota} \over {\partial t}}+{ik\mu\over a}\iota-y_S = \sigma_T n_e\biggl(\sigma_0-4\mu v_b-\iota+P_2(\mu)
1/2(\sigma_2/5-\eta_0+\eta_2/5)\biggr)\cr
{{\partial q} \over {\partial t}}+{ik\mu\over a} q = \sigma_Tn_e\biggl(-q+1/2(1-P_2(\mu))(-\sigma_2/5+\eta_0-\eta_2/5)\biggr)\cr}
\e

These equations are coupled together through the quadrupole term, {\it i.e.} 
the radiation must have a quadrupole anisotropy to get polarized. 

\subsection{Tensor modes}

For tensor perturbations the only non vanishing components of the
perturbed metric tensor are $h_{11}=h_{22}=h_+$ and $h_{12}=h_{21}=h_\times$
where the two values $h_+$, $h_\times$ refer to the two polarization
states of the gravitational waves.
The equation of transfer has the following form [24] :

\i {{\partial \over \partial t}} \pmatrix{\iota\cr q\cr u\cr} + {i k \mu\over a}
\pmatrix{\iota\cr q\cr u\cr} - \pmatrix {  y\cr0\cr0\cr}= 
\sigma_T n_e \biggl( 
  \int_{\Omega} {\hat{M_T}}(\mu,\phi;\mu',\phi') \pmatrix{\iota'\cr
q'\cr u'\cr} {d\Omega' \over 4\pi}
- \pmatrix{{\iota} \cr q \cr u \cr}\biggr)\e 

\

\noindent where now $y = - {\dot{h}_+(1-\mu^2)\cos(2\phi)}+ 
{\dot{h}_\times (1-\mu^{2})\sin(2\phi)}$,
 and  the $3 \times 3$  matrix $\hat{M_T}$ is composed by the
first three rows and columns of the matrix $\hat{P^2}$ in the 
$(I,Q,U,V)$ basis:

\i {\hat{M_T}} = {3\over8}\pmatrix{
K_-(\mu)K_-(\mu')\cos\Delta_\phi&
-K_-(\mu)K_+(\mu')\cos\Delta_\phi&
-2\mu'K_-(\mu)\sin\Delta_\phi\cr
K_+(\mu)K_-(\mu')\cos\Delta_\phi&
-K_+(\mu)K_+(\mu')\cos\Delta_\phi&
2\mu'K_+(\mu)\sin\Delta_\phi\cr
\mu K_-(\mu')\sin\Delta_\phi&
-\mu K_+(\mu')\sin\Delta_\phi&
2\mu\mu'\cos\Delta_\phi\cr} \e

\

\noindent with $K_\pm(\mu)= 1 \pm \mu^2$, $\Delta_\phi = 2(\phi' -\phi)$.
The particular form of the metric tensor
makes $\iota$ still dependent on $\phi$ in spite of the choice of a special
reference system. However, this dependence is not too cumbersome. 
In fact, it is possible to introduce a
change of variables [24] to eliminate the dependence on the azimuthal angle.
The new quantities, $\tilde I$, $\tilde Q$ and $\tilde U$ are related to the old ones by
the following relation:

\i
\cases{I(\mu,\phi)= \tilde I_+(\mu) (1-\mu^2) \cos 2\phi+\tilde I_\times(\mu)
(1-\mu^2)\sin 2\phi\cr
Q(\mu,\phi) = \tilde Q_+(\mu) (1+\mu^2)\cos2\phi+\tilde Q_\times (\mu)
(1+\mu^2) \sin 2\phi\cr
U(\mu,\phi) = -\tilde U_+ 2\mu\sin2\phi+ \tilde U_\times 2\mu\cos2\phi\cr}
\e

\noindent It is easy to prove that, with these new variables,
 only $\hat{P^2}$ provides a non vanishing contribution to the integral
of equation (7) over $\phi$. This is why we have  
considered only this term in equation (21).
Also, as the Boltzmann equation becomes independent of $\phi$, we can
still develop fluctuations of $\tilde I$,$\tilde Q$ and $\tilde U$ in Legendre polynomials.
Thus, equation (20) becomes:

\i
\cases{\dot{\tilde \iota_\epsilon}+{ik\mu\over a} \tilde \iota_\epsilon - 2 \dot
h_\epsilon = - \sigma_T n_e (\tilde \iota_\epsilon+\Psi)\cr
\dot{\tilde q_\epsilon}+{ik\mu\over a} \tilde q_\epsilon=-\sigma_Tn_e(\tilde q_\epsilon-\Psi)\cr
{\tilde q_\epsilon} + {\tilde u_\epsilon} = 0\cr}
\e

\noindent where

\i \Psi= 3/5 \tilde \eta_{\epsilon,0} + 6/35 \tilde \eta_{\epsilon,2} + 1/210
\tilde \eta_{\epsilon,4} - 1/10 {\tilde \sigma}_{\epsilon,0} +1/35
\tilde\sigma_{\epsilon,2}- 1/210\tilde\sigma_{\epsilon,a}
\e 

\noindent and $ \epsilon $ identifies either the $+$ or the $\times$ polarization state of the
gravitational wave.

\subsection{Numerical calculations}

We restrict ourselves to a Universe composed by baryons, cold dark matter, photons and three families of massless
neutrinos. The equations describing anisotropy and polarization of the CMB have been written above.
In Fourier space, the equations describing fractional fluctuations in the remaining cosmic components
are [42,43,44]:

\i
{\partial {\iota}_\nu \over \partial t}+ i{k\mu \over a}\iota_\nu = y
\e

\i
\ddot h + 2{\dot a\over a}\dot h = 8\pi G
\left( \rho_{\scriptscriptstyle B}\delta_{\scriptscriptstyle B}
+\rho_{\scriptscriptstyle CDM}\delta_{\scriptscriptstyle
CDM} + 2 \rho_\gamma\delta_\gamma + 2\rho_\nu\delta_\nu
\right) 
\e

\i
\dot h_{33} - \dot h = {16 \pi G a\over k} \left(
\rho_{\scriptscriptstyle B}v  +
\rho_\gamma f_\gamma + \rho_\nu f_\nu \right) 
\e

\i
\dot \delta_{\scriptscriptstyle B} =
{\dot h\over 2} - i {k v\over a}
\e

\i
\dot v_b + H(t) v_b =
\sigma_{\scriptscriptstyle T} n_e \left( f_\gamma -
{4 \over 3} v_b \right)
\e

\i
\dot \delta_{\scriptscriptstyle CDM} =
{\dot h \over 2}
\e

\noindent and

\i
\ddot h_{+,\times} + 3{\dot a\over a}\dot h_{+,\times} +{k^2 \over a^2} h_{+,\times} = 0
\e

\noindent for scalar and tensor fluctuations, respectively.

Eq.(25) describes the fluctuations in the massless neutrinos.
We follow this component only when the perturbation proper wavelength is larger than one
tenth of the horizon. 
Afterwards, free streaming rapidly damps fluctuations in this hot component.

The time evolution of the baryon and CDM density contrasts and of the baryon
peculiar velocity are described by Eq.(28), (30) and (29) respectively. 
The system for the scalar fluctuations is closed by Eq.(26) and (27) describing the field equations for the trace and
the $3-3$ component of the metric perturbation tensor, while Eq.(31) is all we need to
describe the evolution of the metric perturbations for tensor fluctuations.
 We numerically integrate the previous equations from redshift $z=10^7$ up to
the present. The abundance of free electrons, $n_e$, is evaluated following a 
standard recombination scheme [45,46] for $H$ and $^4He$, 
taken in the ratio $77 : 33$.
 In the following we also consider the possibility that the universe reionized
instantaneously at redshift $z_{rh}<<1000$, and remained completely reionized
up to the present.

\

\section{ Computing the Correlation Function for Anisotropy and Polarization}

\begin{figure}
\centerline{\epsfig{file=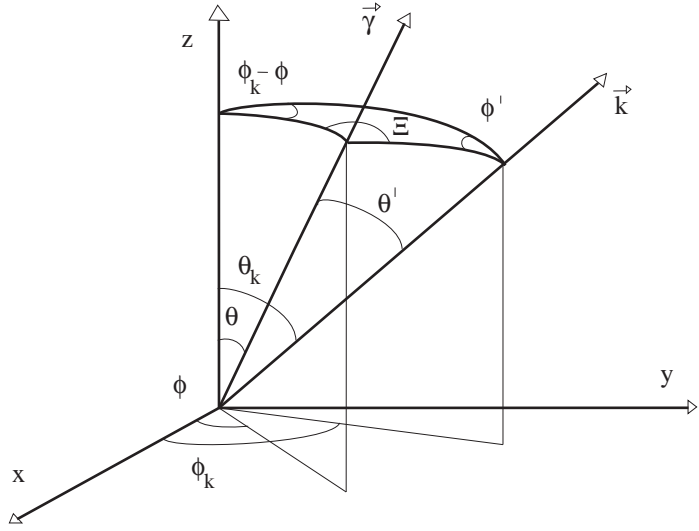,width=2.5in,angle=-90}}
\caption{Laboratory reference system. }
\end{figure}

Under the assumption of gaussian initial conditions, the statistical properties
of the CMB anisotropy and polarization patterns are fully described in terms of their correlation 
functions. The stochastic anisotropic component of the CMB is conveniently expanded in spherical
harmonics: $\delta T (\hat{\gamma})/T_0 = \sum_{lm} a_{lm} Y_m^l (\hat{\gamma})$.
The coefficients $a_{lm}$ are random gaussian variables with zero mean and rotationally invariant
variances, $C_\ell \equiv \langle \mid a_{lm} \mid ^2 \rangle$.
The mean (over the ensemble) correlation function of the anisotropy pattern has the
standard expression:

\i \langle {\delta T(\vec \gamma_1)\over T_0}{\delta T(\vec \gamma_2)\over
T_0}\rangle = {1\over 4\pi} \sum_\ell (2\ell+1) {C_\ell} P_\ell
(\cos\theta)\e

\noindent where $\cos\theta = \vec \gamma_1 \cdot \vec \gamma_2$, and

\i {C_\ell} = {A_S\over 8\pi} \int_0^\infty {\mid\sigma_\ell(k)\mid^2\over
(2\ell+1)^2} k^{n_S+2} dk\e

Here the primordial power spectrum of scalar fluctuations is assumed
to have the standard form $P(k) = A_S k^{n_S}$.
In the case of tensor fluctuations, the change of variables
 needed to achieve rotationally symmetry [see equation (22)] must be taken into account.
Then, the correlation function of the CMB anisotropy induced by tensor
modes reads:

\i
{\langle {{\delta T(\vec \gamma_1)}\over{T_0}} {{\delta T(\vec \gamma_2)}\over{T_0}} 
\rangle} = {{A_T} \over {128 \pi^3}}
 \int {\sum_{\ell 1 \ell 2}  \Pi_{\ell 1, \ell 2}(k,{\vec \gamma'}_1,{\vec \gamma'}_2)
 P_{\ell 1} ({\mu'}_1) P_{\ell 2}({\mu'}_2)
k^{n_T-3} d^3 k }
\e

\noindent where

\i
\Pi_{\ell 1,\ell 2} = K_{-}({\mu'}_1)K_{-}({\mu'}_2) [{{\tilde \sigma}_{\ell 1}^\times}
 {{\tilde \sigma}_{\ell 2}^{\times *}} \cos (2 \phi'_1) \cos (2 \phi'_2)
+{{\tilde \sigma}_{\ell 1}^+} {{\tilde \sigma}_{\ell 2}^{+ *}} \sin (2 \phi'_1) \sin (2 \phi'_2)]
\e
\

\noindent where the power spectrum of metric fluctuations due to tensor modes is assumed to be 
$\tilde P(k) = A_T k^{n_T-3}$. 
Assuming ${\tilde \sigma}_\ell^+={\tilde \sigma}_\ell^\times$ and making
some algebraic manipulations yield:

\i
\langle {{\delta T(\vec \gamma_1)}\over{T_0}}{{\delta T(\vec \gamma_2)}\over{T_0}} 
\rangle ={A_T \over {128 \pi^3}}
\int\sum_{\ell 1 \ell 2} \Upsilon ({\vec \gamma'}_1, {\vec \gamma'}_2) 
{{\tilde \sigma}_{\ell 1}^+} {{\tilde \sigma}_{\ell 2}^{+ *}} 
 P_{\ell 1}({\mu'}_1)P_{\ell 2}({\mu'}_2) k^{n_T-3} d^3 k
\e

\noindent where

\i
\Upsilon ={[2 ({\vec \gamma_1} \cdot {\vec \gamma_2} - \mu'_1 \mu'_2)^2
 -  K_{-}({\mu'}_1)K_{-}({\mu'}_2)]}
\e

\noindent and, finally [47,48,49],

\i \langle {\delta T(\vec \gamma_1)\over T_0}{\delta T(\vec \gamma_2)\over
T_0}\rangle = {1\over 4\pi} \sum_\ell (2\ell+1) {\tilde C_\ell} P_\ell
(\cos\theta)\e

\noindent with 

\i {\tilde C_\ell}= {A_T\over 8\pi} {(\ell+2)!\over(\ell-2)!}\int_0^\infty
{\mid{\tilde \Sigma_{\ell}(k)}\mid^2 \over(2\ell+1)^2}k^{n_T-1}dk\e

\noindent and

\i \tilde \Sigma_{\ell}(k) = {\tilde\sigma^+_{\ell-2}\over (2\ell-1)(2\ell-3)}-
2{\tilde\sigma^+_\ell\over(2\ell-1)(2\ell+3)}+{\tilde\sigma^+_{\ell+2}\over
(2\ell+5)(2\ell+3)}\e

\

As discussed in Section 2, the $Q$ and $U$ Stokes parameters vary because of
 a clockwise rotation $\Xi$ of the reference system in the polarization plane :

\i
\cases{Q_{\Xi}=Q \cos (2 \Xi) + U \sin (2 \Xi )\cr
U_{\Xi} =-Q \sin (2 \Xi) + U \cos (2 \Xi)\cr}
\e

So, the perturbations to $Q$ and $U$ in the Lab frame
are related with those in  $\vec k$ space by 
the following relation:

\i
 {{Q(\vec x, \theta, \phi)} \over T_0}= {1\over 32 \pi^3}\int q(\vec k) e^{i \vec k \vec x} \cos [2 \Xi(\vec k)] d^3 k
\e
\i
 {{U(\vec x, \theta, \phi)} \over T_0}= {{-1}\over 32 \pi^3 }\int q(\vec k) e^{i \vec k \vec x} \sin [2 \Xi(\vec k)] d^3 k
\e

\noindent The correlation function for
the $Q$ parameter for scalar modes can be written as follows:

\i
\langle{Q(\vec \gamma_1)\over T_0}{Q(\vec \gamma_2)\over T_0}\rangle = {A_S \over
{128 \pi ^3}} \int  \sum_{\ell_1,\ell_2} 
\eta_{\ell 1}^* \eta_{\ell 2} \cos (2 \Xi_1) \cos (2 \Xi_2)
P_{\ell 1}(\mu'_1) P_{\ell 2}(\mu'_2) k^{n_S} d^3 k
\e

The correlation function for $U$ has a similar expression with
$\cos 2 \Xi \rightarrow \sin 2 \Xi$.
Let us identify the line of sight $\vec \gamma_1$ with the $z$-axis of the Lab frame.
In this case, $(\theta_1, \phi_1)=(0,0)$, $\theta_1'=\theta_k$ and
$\phi_1'=-\phi_k$ (see Figure 3). In the small angle approximation, $\cos(2 \Xi_1) \sim \cos (2 \Xi_2) \sim \cos (2 \phi_k)$,
and equation (44) yields :

\i
 \langle{Q(\vec z)\over T_0}{Q(\vec \gamma_2)\over T_0}\rangle = A(\theta) + B(\theta, \phi)
\e

\noindent where
\i
A (\theta) = {A_S \over {256 \pi^3}}
 \int \sum_{\ell 1 \ell 2} \eta_{\ell 2} (k) \eta_{\ell 1}^* (k) P_{\ell 1} ({\mu'}_k) P_{\ell 2}({\mu'}_2) 
k^{n_S} d^3 k
\e
\noindent and
\i
B (\theta,\phi) = {A_S \over {256 \pi^3}}
 \int \sum_{\ell 1 \ell 2} \eta_{\ell 2} (k) \eta_{\ell 1}^* (k) P_{\ell 1} ({\mu'}_k) P_{\ell 2}({\mu'}_2) 
\cos(4 \phi_k) k^{n_S} d^3 k
\e
The first term has the standard expression :
\i A(\theta) = {1\over 4\pi} \sum_\ell (2\ell+1) {C_\ell}^Q P_\ell
(\cos\theta)\e
\noindent with
\i {C_\ell}^Q = {A_S\over 16\pi} \int_0^\infty {\mid\eta_\ell(k)\mid^2\over
(2\ell+1)^2} k^{n_S+2} dk\e

For the second term, we can develop $P_{\ell 1} (\cos \theta_k)$ in 
associated Legendre polynomials with $m=4$:
\i
P_{\ell 1}(\cos(\theta_k))=\sum_{\ell'\ge 4}^\infty \alpha_{\ell 1,\ell'}P_{\ell'}^4(\cos(\theta_k))
\e
\noindent where 
\i
\alpha_{\ell 1 \ell'}={{2 \ell' +1}\over{2}} {{(\ell'-4)!} \over {(\ell'+4)!}} A_{\ell 1,\ell'}
\e
\noindent and $A_{\ell 1,\ell'} = \int_{-1}^1 {P_{\ell_1} P_{\ell'}^4 d(\cos \theta_k)}$ has the
following values :

\i
\cases{A_{\ell,\ell}= {2 \over {2 \ell +1}} { \ell ! \over {(\ell - 4)!}}&$\ell' \equiv \ell$\cr
A_{\ell,\ell'}=8(\ell'^2 + \ell' - 3(\ell^2 + \ell +2 ))&$\ell'= \ell+2,\ell+4,...,\ell + 2n$\cr
A_{\ell,\ell'}=0&$\ell'=\ell+1,\ell+3,...,\ell + 2n+1$ \cr
A_{\ell,\ell'} = 0&${\ell'< \ell}$\cr}
\e

From the definition of the spherical harmonics it also follows:

\i
P_{\ell'}^4(\cos \theta_k) \cdot \cos(4 \phi_k) = {1\over2} 
\sqrt{{{4 \pi}\over { 2 \ell' +1}} { {(\ell' +4 )!}
\over {(\ell'-4)!}}} (Y_{\ell'}^4(\vec k) + Y_{\ell'}^{-4} (\vec k) )
\e

\noindent Now, execute the following steps:
\begin{enumerate}

\item  Insert the (50) in the (47)

\item Develop the product between the Legendre polynomial and the cosine with the (53)

\item Integrate in $d \Omega_k$

\item Use again the (53) to transform the spherical harmonics in $P_\ell^4$

\item Develop the $P_\ell^4$ in $P_\ell$. 

\end{enumerate}

\noindent At the end, in the small angle approximation, it is possible to write [29,50,71]:

\i \langle{Q(\vec z)\over T_0}{Q(\vec \gamma_2)\over T_0}\rangle = {1\over 4\pi}
\sum_\ell (2\ell+1) ({C_\ell^Q}+\cos(4\phi)B_\ell) P_\ell (\cos\theta)\e

\noindent where the two terms $C_\ell^Q$ and $B_\ell$ are defined, for scalar
perturbations, as follows:

\i C_\ell^{Q}= {A_S\over 16\pi} \int{\mid\eta_\ell\mid^2\over(2\ell+1)^2}
k^{n_S+2 }dk\e

\i B_\ell={A_S\over 64\pi} \sum_{\ell 1 \ell 2} {{(\ell_2 - 4)!} \over {(\ell_2 + 4)!}}
{A_{\ell , \ell2} A_{\ell1 , \ell2 }} \int {{{\eta_{\ell1}}^*} \eta_{\ell2}
 k^{n_S+2}dk} \e

For tensor fluctuations, the calculation is similar.
The final result is [30,50,71]:

\i \langle{Q(\vec z)\over T_0}{Q(\vec \gamma_2)\over T_0}\rangle = {1\over 4\pi}
\sum_\ell (2\ell+1) ({{\tilde C_\ell}^Q}+\cos(4\phi){ \tilde B_\ell}) P_\ell (\cos\theta)\e

\noindent where

\i\tilde C_\ell^{Q}= {A_T\over 16\pi}
\int{(\mid T_\ell\mid^2 +4 \mid R_\ell\mid^2)\over(2\ell+1)^2}
 k^{n_T-1} dk,\e

\i \tilde B_\ell={A_T\over 64\pi} \sum_{\ell1 \ell2} {(\ell_2 - 4)!\over(\ell_2 + 4)!}
{A_{\ell , \ell2} A_{\ell1 , \ell2 }} \int{{{(T_{\ell1}}^*} T_{\ell2} +
 4R_{\ell1}}^* R_{\ell2})k^{n_T-1} dk,\e

\i R_\ell= {\ell+1\over 2\ell+3} \tilde \eta^+_{\ell+1}+{\ell\over
2\ell-1}\tilde\eta^+_{\ell-1},\e

\i T_\ell= {(\ell+2)(\ell+1)\over(2\ell+3)(2\ell+5)}\tilde\eta^+_{\ell+2}+
2{6\ell^3+9\ell^2-\ell-2\over
(2\ell+3)(2\ell-1)(2\ell+1)}\tilde\eta^+_\ell+{\ell(\ell-1)\over(2\ell-1)
(2\ell-3)}\tilde\eta^+_{\ell-2},\e

\noindent and  $A_{\ell1,\ell2} = \int P_{\ell1} (x) {P_{\ell2}}^4 (x) dx$.

An interesting case is the correlation function between
CMB anisotropy and polarization.
For scalar fluctuations and in the small angle approximation, the result is [29,50,71]:

\i \langle{{\delta T(\vec z)}\over T_0}{Q(\vec \gamma_2)\over T_0}\rangle = {1\over 4\pi}
\sum_\ell (2\ell+1) {C_\ell^{QT}} \cos(2\phi) {P_\ell}^2 (\cos\theta)\e

\noindent where

\i  C_\ell^{QT}={A_S\over 16\pi} \sum_{\ell1} {(\ell - 2)!\over(\ell + 2)!
} {B_{\ell1, \ell}} \int { { { {\sigma_{\ell1}}^*} \eta_\ell} \over
{(2\ell+1)}} k^{n_S+2} dk\e

\noindent and  the integral $B_{\ell1,\ell2} = \int P_{\ell1} (x) {P_{\ell2}}^2 (x) dx$ has the values :

\i
\cases{B_{\ell,\ell}= - {2 \over {2 \ell +1}} { \ell ! \over {(\ell - 2)!}}&$\ell'=\ell$ \cr
B_{\ell,\ell'}= 4 &$\ell'=\ell+2,\ell+4,...,\ell+2n$ \cr
B_{\ell,\ell'}=0 &$ \ell'=\ell+1,\ell+3,...,\ell+2n+1$ \cr
B_{\ell,\ell'} = 0 &$ \ell'< \ell$\cr}
\e

\section{ Numerical Results and Discussion}

With the formalism developed in the previous Sections, we are now able
to make theoretical predictions for CMB anisotropy and polarization.
As stated before, we restrict ourselves to the Cold Dark Matter scenario. This is 
not quite enough to completely define the model, as we have to deal
with quite a number of parameters. We
have to fix : i) the total density parameter, $\Omega_0$, and the cosmological
constant, $\Lambda$; ii) the baryonic abundance, $\Omega_b$; iii) the Hubble constant;
iv) the primordial spectral index for spectral fluctuations;
v) the relative amplitude of scalar and tensor fluctuations;
vi) the spectral index for tensor fluctuations; vii) the thermal history of the
universe; viii) the overall amplitude for scalar fluctuations.

The baryonic abundance is quite severly restricted by primordial
nucleosynthesis. We consider the fiducial value of $\Omega_b = 0.05 \pm 0.02$ as
representative of the possible uncertainty in this parameter.
We remind that changes in $\Omega_b$ yield variations in the pressure of the
photon-baryon fluid before recombination, and then variations in
 the amplitude of the first acoustic peak in the anisotropy
power spectrum.

For flat models the value of the Hubble constant
is usually taken to be $H_0 = 50 {Km}  {s^{-1}} / Mpc$ for age considerations,
 even if the estimated globular cluster age allows for slightly 
different values: $ 40 {Km}  {s^{-1}} / Mpc < H_0 < 65 {Km}  {s^{-1}} / Mpc$ [63]. Small
variations in the Hubble constant yield huge variations in the
radiation power spectrum, modifying both the amplitudes and positions of 
the acoustic peaks. 

The primordial index for scalar fluctuations,
$n_S$, is usually taken to be unity, as inflation suggests.
However, in more general inflationary scenarios, $n_S$ can be
either smaller or larger than unity [64]. This modifies the
relative power between the anisotropy on small and large angular scale.
Power-law inflationary models predict $n_S < 1$, but also a background
of gravitational waves. A standard prediction is that the ratio
of the quadrupoles induced by scalar and tensor fluctuations is:

\i
{{\tilde C}_2 \over C_2} = -7n_T \sim 7(1-n_s),
\e

\noindent which allows to relate amplitudes and shapes of primordial 
power spectra for scalar and tensor fluctuations, respectively [18,49,64].

The thermal history of the universe, in its standard form, assumes 
recombination of the primordial plasma at redshift $\sim 1000$. However,
both the Gunn-Peterson test [65] and the enriched composition of the
intracluster medium [74,75] suggest the possibility of a considerable
 energy release during the early stages of galaxy formation and evolution.

 Finally, the amplitude of fluctuations it is still unknown from first
principles, and it is fixed in order to match the observed rms
temperature fluctuations ( $29 \pm 1 \mu K$) of the COBE/DMR anisotropy maps [53].

\begin{figure}
\centerline{\epsfig{file=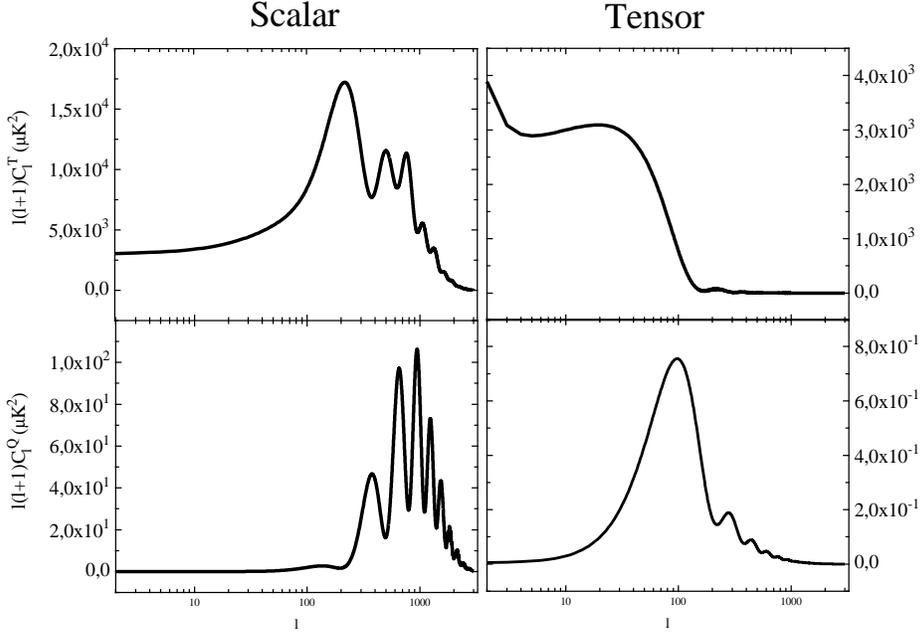,width=3.8in,angle=-90}}
\caption{Anisotropy (top) and polarization (bottom) power spectrum for
scalar (left) and tensor (right) fluctuations.Each model is normalized
to COBE $\sigma (10^o) = 29 \mu K$.}
\end{figure}

In Figure 4 we show theoretical predictions for CMB anisotropy and 
polarization of a standard Cold Dark Matter Model with $\Omega_0 =1$,
$\Omega_b=0.05$, $H_0 = 50 {Km} {s^{-1}} / Mpc$, $n_S= 1$ and standard recombination. The
anisotropy power spectrum for scalar fluctuations has a flat
behavior at low $\ell$'s, where the Sachs-Wolfe effect [66] dominates,
and a structure of peaks at higher $\ell$'s, defined by the
acoustic oscillations in the photon-baryon fluid experienced before
 recombination by fluctuations smaller then the acoustic horizon.
 The damping at high $\ell$'s is due to the finite thickness of
the last-scattering surface [67]. The first peak at $\ell \sim 200$
corresponds to fluctuations that entered the horizon at recombination,
 which subtends an angle $\sim 2^o h^{-1}$. The polarization power
 spectrum has instead power only at $\ell >200$, {\it i.e.} on
scales $< 2^o h^{-1}$.
The case of pure tensor fluctuations is show in Figure 4 only for
the didactic purposes. In this case, the anisotropy spectrum has power
 only at $\ell < 200$ and the polarization spectrum shows a prominent
peak at $\ell \sim 100$. Note that the polarization spectrum has an
amplitude lower than the anisotropy spectrum by a factor $\sim 10^2$
 and $\sim 10^4$ for scalar and tensor fluctuations, respectively.

Real experiments are sensitive to a limited region of the power spectrum, 
because of the antenna beam and modulation techniques.
For anisotropy experiments, this effect can be parameterized by a window  
function, $W_\ell$, so that the variance of temperature fluctuations detected by an experiment
can be written as:

\i \langle {\biggl({ {\delta T(\vec \gamma)} \over T_0} \biggr)}^2 \rangle 
= {1\over 4\pi} \sum_\ell (2\ell+1) C_\ell W_\ell\e

A similar expression holds for the variance of the fluctuations of the $Q$
 parameter:

\i \langle {\biggl({Q(\vec \gamma)\over T_0} \biggr)}^2 \rangle = {1\over 4\pi} \sum_\ell (2\ell+1) {C_\ell}^Q W_\ell \e

\noindent In fact, it can be proved that the azimuthal contribution [see equations (54) and (57)] 
to the $Q$ variance
vanishes. In equation (67) we use the same anisotropy window functions, in order
to give an order of magnitude estimate of the level of measurable polarization
at different angular scales.

\begin{figure}
\centerline{\epsfig{file=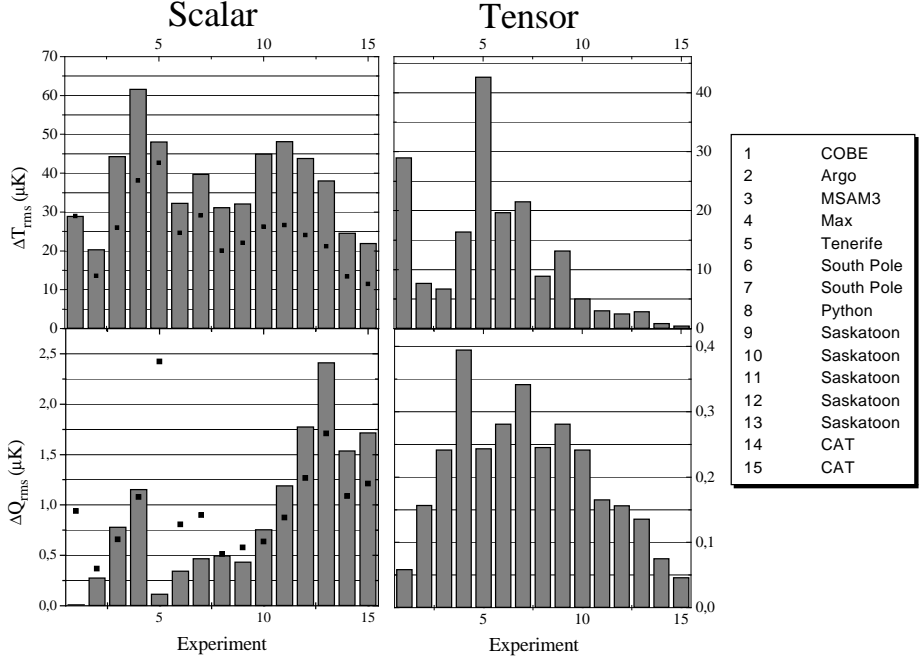,width=3.8in,angle=-90}}
\caption{R.m.s. values for anisotropies (top) and polarization (bottom)
scheduled for scalar (left) and tensor (right) models for different
experiments [53...61].The tensor model has $n_T=0$ and the scalar model has $n_S=1$.
Each model is normalized to COBE $\sigma (10^o) = 29 \mu K$. 
The dot of the left side represents the values for a scalar reionized model
at $z_r \sim 70$.}
\end{figure}

In Figure 5 we plot the expected rms values for CMB anisotropy and polarization
using 15 different window functions corresponding to 9 different anisotropy
experiments.
The level of polarization from scalar modes is below the current experimental sensitivity,
even for small scale experiments such has Saskatoon or CAT that are sensitives
to multipole $\ell \sim 400$ where the polarization has the first two peaks.
The rms level from pure tensor modes, even for the MAX experiment that seems
to have the best window function, is below 0.5 $\mu K$, so the
separation between scalar and tensor fluctuations do not seems to be at hand with
polarization measurements [25,26,27,28,73].
This can be done by  combining anisotropy measurements
at both large (where tensor modes could contribute) and small (where tensor modes 
do not contribute) angular scales [49,72,73]. An accurate mapping of the anisotropy pattern 
with both high sensitivity and high angular resolution will be provided by planned, 
dedicated space missions such as COBRAS/SAMBA [68] and MAP [52]. At the moment
the bulk of degree-scale detections, combined with the COBE/DMR and Tenerife experiments,
 seems to suggest a spectral index for scalar fluctuations $n_S \ge 1$ [69] and
a negligible contribution of tensor modes.

\begin{figure}
\centerline{\epsfig{file=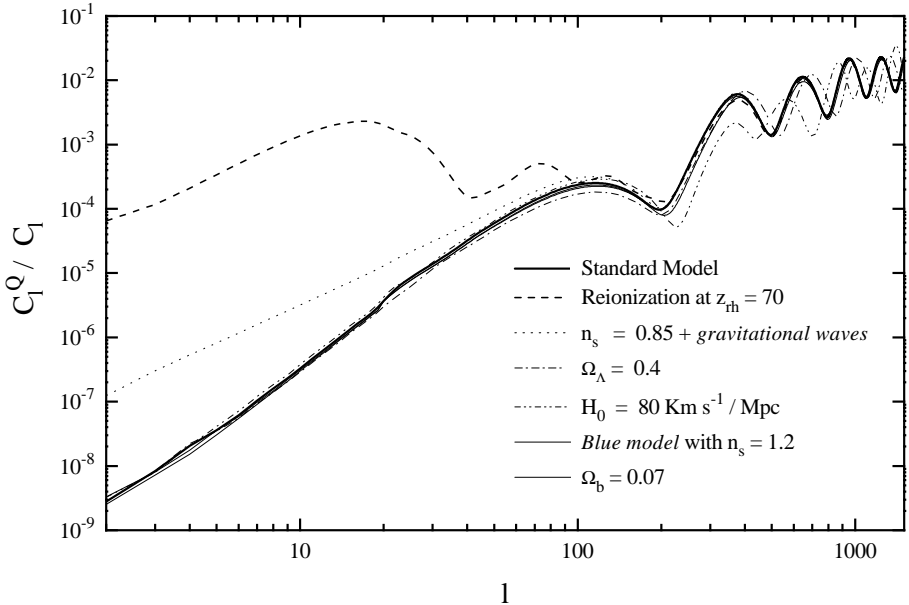,width=3.8in,angle=-90}}
\caption{Dependence of polarization on cosmological parameters.}
\end{figure}

As mentioned above, there are several free parameters, each with its own uncertainty,
 which define a theoretical model. So, it is interesting to explore the sensitivity of 
the polarization level relative to the anisotropy one. To show this, we plot in
 Figure 6 the quantity $C_\ell^Q / C_\ell$ as a function of $\ell$, for different
choices of the model parameters. Generally speaking the effects of the variation of these parameters on
anisotropy and polarization are similar: both quantities tend to decrease with
increasing $H_0$ and tend to increase when a cosmological constant $\Omega_\Lambda = 1-\Omega_0$
is taken into account. In particular, varying $n_S$ or $\Omega_b$ yields 
basically no variations on the $C_\ell^Q / C_\ell$ ratio.
Moreover, decreasing the spectral index and adding gravitational waves increase 
the large scale polarization, but, as we have shown, not enough to pass
the threshold of present day detector sensitivity.
So, even taking into account reasonable uncertainties in the parameters,
its seems that only with an huge increment in the experimental sensitivity
 (see the accompanying paper by F. Melchiorri et al. in this volume) and/or a  space mission [52,70,71]
 a robust detection of the 
polarization spectrum over a wide range of $\ell$'s would be possible.
Coulson et al. [29] suggested searching  a correlation between the temperature
in one direction and the polarization in a circle at distance $\Theta$ from
that direction. The shape of the correlation function (62) measurable in this
way is show in Figure 7. As we can see the cross correlation is positive on
scales $>1^o$, negative on scales between $0.5^o$ and $1^0$ and positive again on
scales $<0.5^o$. According to [52] the future MAP satellite would have the
capability to measure the expected amplitude of this signal.

\begin{figure}
\centerline{\epsfig{file=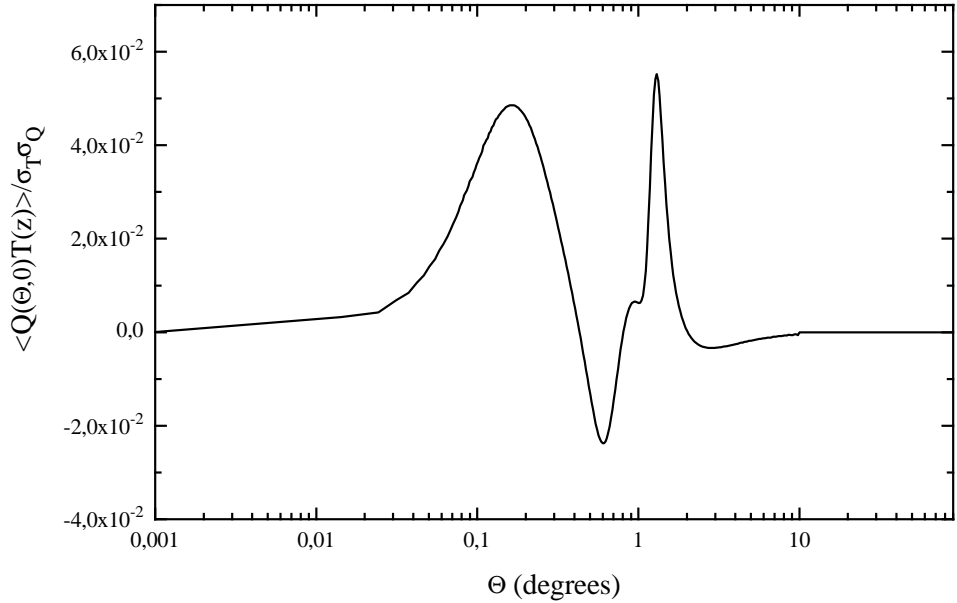,width=3.8in,angle=-90}}

\caption{Polarization - Anisotropy correlation function.}
\end{figure}

The final item to be investigated is the dependence of CMB anisotropy and 
polarization on the thermal history of
the universe. A reionization at $z \le 100$ produces a new, later and thicker 
last scattering surface.
The effect of such a new last scattering surface is to smooth the
anisotropy on small angular scales and to leave unchanged the level of anisotropy on
large angular scales. The effect of reionization on polarization is to reduce the
polarization on small scales but to increase the polarization level at large angular
scales. This is shown again in Figure 6. Thus, possible detection of polarization
 between $ 1^o$ and $10^o$, say, would be an evidence for an early re-heating of the 
intergalactic medium.

\section{Conclusions}

Numerical solutions of the Boltzmann
transport equation show that a certain degree of polarization must be present
as a consequence of the primordial fluctuations responsible for the structure formation.
The level of polarization depends strongly on the angular scale,
much more than in the case of anisotropy, quite independently of the choice of the model
parameters. At angular scales larger than one
degree we do not expect detectable polarization unless the universe was
reionized at early times $z\geq 40$. At small angular scales the polarization
may reach the $5-10 \%$ of the anisotropy. However, a polarization of a few
 percent at angular scales
of $1^o$-$10^o$ could be explained only because  reionization: a search for
polarization at these  scale is therefore important in the study of the
thermal history of the universe.
Also, it seems hard to disentangle tensor 
from scalar perturbations through measurements of polarization, due to the tenuity of the
signal.

\section{Aknowledgments}

We would like to thank Paolo de Bernardis for comments and suggestions.
AM thanks Arthur Kosowsky for helpful discussions.
This work has been supported by MURST.

\end{document}